\documentclass[aps,pra,twocolumn,superscriptaddress,notitlepage,amsmath,amssymb,floatfix]{revtex4-2}

\usepackage{amsmath,amsthm,amssymb,bm,graphicx,float,xcolor,changes,mathtools,enumitem}
\usepackage[colorlinks=true,urlcolor=blue,linkcolor=blue,citecolor=green,bookmarks=false]{hyperref}

\allowdisplaybreaks

\newcommand{\be}{\begin{equation}}
\newcommand{\ee}{\end{equation}}
\newcommand{\6}{\partial}

\newcommand{\bk}{\boldsymbol{k}}
\newcommand{\bl}{\boldsymbol{\lambda}}
\newcommand{\bx}{\boldsymbol{x}}
\newcommand{\bs}{\boldsymbol{\sigma}}
\DeclareMathOperator{\sech}{sech}

\begin{document}

\title{Expansion of one-dimensional spinor gases from power-law traps}

\begin{abstract}
Free expansion following the removal of axial confinement represents a fundamental nonequilibrium scenario in the study of many-body ultracold 
gases. Using the stationary phase approximation, we analytically demonstrate that for all one-dimensional spinor gases with repulsive contact 
interactions, whether bosonic or fermionic, the asymptotic density and momentum distribution can be directly determined from the 
quasimomentum distribution (Bethe rapidities) of the trapped gas. We efficiently obtain the quasimomentum distribution numerically by solving 
the integral equations that characterize the ground state of the integrable system within the local density approximation. Additionally, we 
derive analytical solutions for both weakly and strongly interacting regimes.
Unlike in bosonic gases, where rapidity distributions and density profiles vary significantly across interaction regimes, fermionic gases 
maintain similar profiles in both weakly and strongly interacting limits. Notably, the gas expands self-similarly only when released from a 
harmonic trap. For other power-law trapping potentials, the asymptotic density profile is strongly influenced by the initial confinement 
geometry. Our results extend readily to Bose-Fermi mixtures and finite temperatures.
\end{abstract}

\author{Ovidiu I. P\^{a}\c{t}u}
\affiliation{Institute for Space Sciences, Bucharest-M\u{a}gurele, R 077125, Romania}
\author{Gianni Aupetit-Diallo}
\affiliation{SISSA, Via Bonomea 265, I-34136 Trieste, Italy}

\maketitle

\section{Introduction}

The nonequilibrium dynamics of many-body quantum systems presents exceptional challenges, both numerically and analytically. Despite 
significant  progress in recent decades through analytical techniques such as hydrodynamical approaches \cite{CD96,KSS96,KSS97,MS02,PSOS03,
FCJB14}, the Quench Action \cite{CE13,Caux16}, and, more recently, Generalized Hydrodynamics (GHD) \cite{CDY16,BCNF16,BD22,MZLD21}, our understanding of 
far-from-equilibrium dynamics remains incomplete.

A common nonequilibrium scenario in many-body physics is the free expansion of a system following a sudden quench of the trapping potential. In 
many experimentally relevant cases, the system's post-quench evolution is governed by an integrable one-dimensional (1D) Hamiltonian \cite{KBI93}. 
Early analytical studies  focused on the harmonically trapped Lieb-Liniger  model \cite{LL63} with strong repulsive interactions, revealing that after release, 
the asymptotic momentum distribution approaches that of a system of free fermions in the initial trap, a phenomenon known as dynamical 
fermionization \cite{RM05,MG05} (see also \cite{delC08,GP08,BHLM12,WR17}), which was recently confirmed experimentally \cite{WMLZ20}. Subsequent 
studies at finite interaction strengths revealed that the asymptotic momentum distribution after the quench coincides with the Bethe rapidity 
distribution of the trapped gas \cite{CGK15,CDDK19}, previously thought to be unobservable.
In multicomponent systems, the presence of the spin sector in the Bethe ansatz wavefunctions significantly complicates analytical investigations, 
even in equilibrium. This complexity is further exacerbated in nonequilibrium settings, making the dynamics of spinor gases particularly 
challenging to study \cite{EN17,MBPC17,SMS18,ZVR19,WYCZ20,NT20,MKKB20,NT21,SCP21,RBC22,SCP22,BFZ19,PVM22,MVBF23,MAMV24,MAVM25}. Only recently has an 
analytical proof of dynamical fermionization been established for multicomponent systems in the Tonks-Girardeau regime, both at zero 
\cite{ASYP21,Patu22} and finite temperature \cite{Patu23}.

In this article, we investigate the asymptotic momentum distribution and density profile of 1D multicomponent spinor gases with short-range 
repulsive interactions, both bosonic and fermionic, after release from a power-law trapping potential. We demonstrate that both quantities can be 
determined from the quasimomentum distribution of the trapped gas, which can be efficiently computed using the integral equations for the ground 
state of the integrable system \cite{Taka99} combined with the local density approximation. For strong- and weak-interaction regimes, we derive 
analytical solutions for the initial density profile and quasimomentum distributions. We show that while bosonic systems are strongly influenced by 
interaction strength, in fermionic systems, the weak- and strong-interaction results coincide up to a multiplicative factor equal to the number of 
components. Furthermore, we prove that self-similar solutions for the density profile emerge only after release from a harmonic potential, whereas 
for all other power-law exponents, the density profile evolution is significantly affected by trap anharmonicity.
Our results extend the findings of Campbell, Gangardt, and Kheruntsyan \cite{CGK15} for the bosonic Lieb-Liniger model in a similar nonequilibrium scenario to 
multicomponent systems with an arbitrary number of components and general quantum statistics. The analysis of such systems presents significant challenges due 
to the highly intricate structure of the wavefunctions, which now include contributions from the spin sector. Additionally, the absence of an explicit expression 
for the normalization factor, crucial in the original derivation for the Lieb-Liniger model, complicates the analysis further. In the main text, we demonstrate 
how these difficulties can be effectively addressed.

\section{Model and quench protocol}

We consider a $\kappa$-component system of one-dimensional particles with contact interactions in the presence of a power-law confining potential.
At $t=0$ the external potential is removed, and our main goal is to investigate the asymptotic density and momentum distribution following the 
expansion. In the second quantization formalism, the Hamiltonian is given by
\begin{align}\label{ham}
H=&\int dx\, \frac{\hbar^2}{2m}\left(\6_x \boldsymbol{\Psi}^\dagger\6_x \boldsymbol{\Psi}^\dagger\right)+
\frac{\mathfrak{g}}{2}:\left(\boldsymbol{\Psi}^\dagger\boldsymbol{\Psi}\right)^2:\nonumber\\
&\qquad\qquad\qquad +V(x,t)\boldsymbol{\Psi}^\dagger\boldsymbol{\Psi}-\boldsymbol{\Psi}^\dagger\boldsymbol{\mu}\boldsymbol{\Psi}\, ,
\end{align}
where $m$ is the particle mass, $\mathfrak{g}>0$  is the strength of the repulsive interaction and  $\boldsymbol{\Psi}=\left(\Psi_1(x),\cdots,
\Psi_\kappa(x)\right)^T$, $\boldsymbol{\Psi}^\dagger =\left(\Psi_1^\dagger(x),\cdots,\Psi_\kappa^\dagger(x)\right)$ with $\Psi_\sigma(x)$, 
$\sigma\in\{1,\cdots,\kappa\}$   being fermionic or bosonic field operators. These operators satisfy the commutation or anticommutation 
relations $\Psi_\sigma(x)\Psi_{\sigma'}^\dagger(y)-\epsilon \Psi_{\sigma'}^\dagger(y)\Psi_\sigma(x)=\delta_{\sigma.\sigma'}\delta(x-y),$ where 
$\epsilon=-1$ for fermions and $\epsilon=1$ for bosons.
In the Hamiltonian (\ref{ham}) $\boldsymbol{\mu}$  is a diagonal matrix with elements $(\mu_1,\cdots,\mu_\kappa)$ representing the chemical 
potentials of each component, and $:\ \ :$ denotes normal ordering. The external potential implementing the quantum quench has a power-law form, 
$V(x,t)=\Theta(-t)\alpha_\nu|x|^\nu/2, $ where $\Theta(t)$ is the Heaviside function and $\nu>0$.

We denote the initial Hamiltonian (before the quench) as $H_I=H(t<0)$  and the final Hamiltonian (governing the time evolution after the quench) 
as $H_F=H(t>0)$ While the eigenfunctions of $H_I$ can be explicitly obtained only in the limiting cases $\mathfrak{g}=0$ and $\mathfrak{g}=\infty$, 
the final Hamiltonian  $H_F$ remains integrable for all values of the interaction strength. Specifically, it corresponds to: the Lieb-Liniger model 
\cite{LL63} for $\kappa=1$, the Gaudin-Yang model \cite{Gaud67,Yang67} for $\kappa=2$ and  the general Sutherland models \cite{Suth68} for 
$\kappa\ge 3$.

The eigenstates of $H_F$, denoted as  $|\boldsymbol{\psi}(\bk,[\bl])\rangle$, are described by $\kappa$ sets of parameters: the quasimomenta 
$\bk=\{k_i\}_{i=1}^{N}$ and the additional spin rapidities $[\bl]=\left(\{\lambda_i^{(1)}\}_{i=1}^{N_1},\cdots,
\{\lambda_i^{(\kappa-1)}\}_{i=1}^{N_{\kappa-1}}\right)$ where the integers satisfy $N\ge N_1\ge\cdots\ge N_{\kappa-1}$. Here  $N$ represents the 
total number of particles in the system. The number of particles in the state labeled by $\sigma$ is given by $M_\sigma=N_{\sigma-1}-N_\sigma$ 
where, by convention,  $N_0=N$ and $N_\kappa=0$. Since we consider only systems with repulsive interactions, the quasimomenta $k_i$ are all real, 
while the spin rapidities $\lambda_i^{(\sigma)}$  can, in general, be complex (see Chap.~12 of \cite{Taka99}). Moreover, we assume a specific 
ordering of these parameters.  
Within a given sector characterized by $\bold{M}=[M_1,\cdots,M_\kappa]$  the eigenstates of $H_F$  satisfy the energy eigenvalue equation  
$ H_F |\boldsymbol{\psi}(\bk,[\bl])\rangle=\sum_{j=1}^N \frac{\hbar^2 k_j^2}{2 m} |\boldsymbol{\psi}(\bk,[\bl])\rangle$. These states are assumed  
to be orthonormal and form a complete basis meaning they satisfy: 
a) $ \langle\boldsymbol{\psi}(\bk',[\bl']) |\boldsymbol{\psi}(\bk,[\bl])\rangle=\delta_{\bk,\bk'} \delta_{[\bl],[\bl']}$
and b) $\boldsymbol{1}=\int^{<} d\bk\sum_{[\bl]}^{<}|\boldsymbol{\psi}(\bk,[\bl])\rangle \langle \boldsymbol{\psi}(\bk,[\bl])|$. In the 
completeness  relation the integrals and sums are defined as: $\int^{<} d\bk=\int_{k_1<\cdots<k_N} dk_1\cdots dk_N$ and 
$\sum_{[\bl]}^{<}=\sum_{\lambda_1^{(1)}<\cdots<\lambda_{N_1}^{(1)}}\cdots \sum_{\lambda_1^{(\kappa-1)}<\cdots<
\lambda_{N_{\kappa-1}}^{(\kappa-1)}}.$

For $N$ large consider the time evolution of the initial state $|\boldsymbol{\phi}_0\rangle$ which is the ground state of the trapped Hamiltonian 
$H_I$. Expanding it in terms of the eigenstates of $H_F$, the time evolution after the quench is given by $|\boldsymbol{\phi}_0(t)\rangle=
e^{-i H_F t/\hbar}|\boldsymbol{\phi}_0\rangle$ or, explicitly
\be\label{e1}
|\boldsymbol{\phi}_0(t)\rangle=\int^{<} d\bk\sum_{[\bl]}^{<} b(\bk, [\bl])e^{- i \sum_j \frac{\hbar k_j^2 t}{2 m} }|\boldsymbol{\psi}(\bk,[\bl])
\rangle
\ee
where the expansion coefficients are given by $ b(\bk, [\bl])=\langle \boldsymbol{\psi}(\bk,[\bl])|\boldsymbol{\phi}_0\rangle$. Assuming the initial 
state normalized, $\langle\boldsymbol{\phi}_0|\boldsymbol{\phi}_0\rangle=1$, this implies the normalization condition $\int^{<} d\bk\sum_{[\bl]}^{<} 
|b(\bk, [\bl])|^2=1$.
The primary quantities of interest are the real space density 
\be\label{density}
\rho(x,t)=\sum_{\sigma=1}^\kappa\langle \boldsymbol{\phi}_0(t)|\Psi_\sigma^\dagger(x)\Psi_\sigma(x)|\boldsymbol{\phi}_0(t)\rangle\, ,
\ee
and the momentum distribution function
\be\label{momentum}
n(p,t)=\sum_{\sigma=1}^\kappa  \int \frac{e^{i p(x-y)}}{2\pi} \langle\boldsymbol{\phi}_0(t)|\Psi_\sigma^\dagger(x)
\Psi_\sigma(y)|\boldsymbol{\phi}_0(t)\rangle\, dxdy\, \, .
\ee
A key observation in deriving the asymptotic distributions of these quantities is that, in Eq.~(\ref{e1}), time appears solely in the exponential 
term, which exhibits rapid oscillations for large $t$ \cite{CGK15}. Combined with the specific structure of the Bethe ansatz wavefunctions, this 
allows us to determine the large-time asymptotics after the trap release using the stationary phase approximation.

\section{The two-component case}

We begin with two-component systems, as their treatment already contains the necessary ingredients required for the general case analysis 
(regarding the single-component Lieb-Liniger model, see Ref.~\cite{CGK15}). For the two-component system, also known as the Gaudin-Yang model 
\cite{Gaud67,Yang67}, the eigenstates in the sector with a total number of particles  $N$ where $M$ particles have spin-down and $N-M$ have spin-up, 
are given by
\begin{widetext}
\be\label{e2}
|\boldsymbol{\psi}(\bk,\bl)\rangle=\frac{1}{\sqrt{N!}}\int dx_1\cdots dx_N\sum_{\sigma_1,\cdots,\sigma_N=\{\uparrow,\downarrow\}}^{[N-M,M]}
\psi(\bx;\bs|\bk,\bl) \Psi_{\sigma_N}^\dagger(x_N)\cdots\Psi_{\sigma_1}^\dagger(x_1)|0\rangle\, ,
\ee
with the wavefunctions \cite{POK12,EFGKK05}
\be\label{e3}
\psi(\bx;\bs|\bk,\bl)=\frac{1}{C(\bk,\bl|\epsilon)}\sum_{Q\in S_N}\Theta(Q\bx)(\epsilon)^Q\sum_{P\in S_N}(-1)^P\varphi(P\bk|\epsilon)
A(Q\bs|P\bk,\bl)e^{i(P\bk,Q\bx)}\, .
\ee
\end{widetext}
We now explain the notation.  
First in Eq.~(\ref{e2}) the notation $[N-M,M]$ above the summation sign indicates that the sum runs over the $C^N_M$ possible combinations in which 
$N-M$ creation  operators correspond to spin-up states and $M$ correspond to spin-down states.  In Eqs.~(\ref{e2}) and (\ref{e3}) we define 
$\bx=(x_1,\cdots,x_N)$ and $S_N$ is the group of permutations of $N$ elements. The term $(-1)^Q$ denotes the signature  of the  permutation $Q$ 
while $Q\bx=(x_{Q_1},\cdots,x_{Q_N})$. The function  $\Theta(Q\bx)=\Theta(x_{Q_1}<\cdots<x_{Q_N})$ is a generalized Heaviside function which takes 
the value $1$ when $x_{Q_1}<\cdots<x_{Q_N}$ and zero otherwise. The notation  $(P\bk,Q\bx)$ represents the sum $\sum_{j=1}^N k_{P_j} x_{Q_j}$ and
the function $\varphi(P \bk|\epsilon)$ is defined as $\varphi(P \bk|\epsilon) =\prod_{ m < n } (k_{P_m} - k_{P_n} + i c)$ for $\epsilon = 1$ and 
$\varphi(P \bk|\epsilon)=1$ for $\epsilon = -1$, where $c=m \mathfrak{g}/\hbar^2$.
The normalization constant $C(\bk,\bl|\epsilon)$ is unknown.  Note that in the Lieb-Liniger case, the normalization constant is known both for finite 
\cite{Kore82,Dorl93} and infinite systems \cite{Gaud71a,Gaud71b},  whereas for multicomponent systems, only a conjecture exists for the finite-size 
Hubbard  model \cite{GK99}. However, on general grounds, $C(\bk,\bl|\epsilon)$ should be symmetric in both the quasimomenta $k$'s and the rapidities 
$\lambda$'s. That is $C(V\bk,\bl|\epsilon)=C(\bk,\bl|\epsilon)$ for all $V\in S_N$ and $C(\bk,V'\bl|\epsilon)=C(\bk,\bl|\epsilon)$ for 
all  $V'\in S_M$. The amplitudes $A(Q\bs|P\bk,\bl)$ are given by as
\begin{align}
A(Q\bs|P\bk,\bl)&=\sum_{R\in S_M}\prod_{1 \le m<n\le M}\frac{\lambda_{R_m}-\lambda_{R_n}+i\epsilon c}{\lambda_{R_m}-\lambda_{R_n}}\nonumber \\
&\qquad\qquad\qquad \times \prod_{l=1}^MF_{P\bk}(\lambda_{R_l};y_l)\, ,
\end{align}
where
\begin{align}
F_{\bk}(\lambda;y)=\frac{i\epsilon c}{\lambda-k_y-i\epsilon c/2}\prod_{j=1}^{y-1}\frac{\lambda-k_j+i\epsilon c/2}{\lambda-k_j-i\epsilon c/2}\, ,
\end{align}
and $y_j$ denotes the position of the $j$-th down spin in $(\sigma_{Q_1},\cdots, \sigma_{Q_N})$.

An important observation is that  $A(Q\bs|P\bk,\bl)$ represents the coordinate Bethe ansatz wavefunctions of an integrable inhomogeneous $XXX$ spin 
chain with inhomogeneities $P\bk=(k_{P_1},\cdots,k_{P_N})$ (see Appendix 3.~B of \cite{EFGKK05}). Consequently, the eigenvectors of this 
inhomogeneous spin chain form an orthogonal and complete basis. Taking into account that for an arbitrary permutation $Q\in S_N$ the summation over 
$\sigma$'s in Eq.~(\ref{e2}) is equivalent to summing over  $\sum_{1\le y_1<\cdots<y_M\le N}$ we obtain 
\begin{align}\label{e5}
\sum_{\sigma_1,\cdots,\sigma_N=\{\uparrow,\downarrow\}}^{[N-M,M]}
A^*(Q\bs|P\bk,\bl')A(Q\bs|P\bk,\bl)=\delta_{\bl,\bl'} D(\bk,\bl)
\end{align}
where $D(\bk,\bl)$ is the normalization constant, which, from general principles, should be symmetric in both the $k$'s and $\lambda$'s 
\cite{Kore82,GK99}.

The wavefunctions in Eq.~(\ref{e3}) are assumed to form a complete orthonormal basis and satisfy the symmetry properties $\psi(\bx;\bs|V\bk,\bl)=
(-1)^V \psi(\bx;\bs|\bk,\bl)$ and $\psi(\bx;\bs|\bk,V'\bl)= \psi(\bx;\bs|\bk,\bl)$ for any $V\in S_N$ and $V'\in S_M$. Consequently, the expansion 
coefficients appearing in Eq.~(\ref{e1}) obey $b(V\bk,\bl)=(-1)^V b(\bk,\bl)$ and $b(\bk,V'\bl)= b(\bk,\bl)$ for any $V\in S_N$ and $V'\in S_M$. 
An important quantity in our asymptotic analysis is the quasimomentum distribution of the trapped gas, defined by
\be\label{defg}
g(k)=N\int dk_2\cdots dk_N\sum_{\lambda_1}\cdots\sum_{\lambda_M}\frac{|b(k,k_2,\cdots,k_N,\bl)|^2}{N!M!}\, ,
\ee
with the normalization $\int g(k)\, dk=N$.

Now, we are ready to perform the asymptotic analysis. Substituting Eq.(\ref{e2}) into Eq.(\ref{e1}), we observe that in the large-$t$ limit, the 
dominant contributions arise  from the stationary phase analysis \cite{BH86} of integrals of the form $\int h(\bk,\bl) e^{i t \left(\frac{kx}{t}-
\frac{\hbar k^2}{2m}\right)}\, dk$ where  $h(\bk,\bl)$ is a slowly varying function of $\boldsymbol{k}$. For each such integral, the stationary 
point is given by  $k_0=m x/t\hbar$. Carrying out this analysis (see Appendix~\ref{as1}) we obtain the asymptotic wavefunction (for the Lieb-Liniger model see \cite{JPGB08,BPG08,CGK15})
\begin{align}\label{e6}
\psi_0(\bx;\bs|t)=&\frac{1}{M!}\left(\frac{2\pi m}{t\hbar}\right)^{\frac{N}{2}}\sum_{\lambda_1}\cdots\sum_{\lambda_{M}}\sum_{Q\in S_N}\Theta(Q\bx)
\nonumber\\
& \ \ \times\frac{b(Q\tilde{\bx},\bl)}{C(Q\tilde{\bx},\bl|\epsilon)}\varphi(Q\tilde{\bx}|\epsilon)A(Q\bs|Q\tilde{\bx},\bl)\nonumber\\
&\ \  \times e^{-i N\pi/4}e^{i\sum_{j=1}^N m x_{Q_j}^2/2m}\, ,
\end{align}
where $Q\tilde{\bx}=\left(\frac{ m x_{Q_1}}{t\hbar}, \cdots, \frac{ m x_{Q_N}}{t\hbar}\right)$. 
Since we assume that the initial wavefunction is normalized to one, the asymptotic wavefunction (\ref{e6}) must also be normalized. Together with 
the normalization condition  on the expansion coefficients, $\int^{<} d\bk\sum_{[\bl]}^{<} |b(\bk, \bl)|^2=1$ this implies that $\varphi 
(\bk|\epsilon)$, $D(\bk,\bl)$ and $C(\bk,\bl|\epsilon)$ satisfy
\begin{align}
(2\pi)^N \frac{N!}{M!}\frac{|\varphi (\bk|\epsilon)|^2 |D(\bk,\bl)|^2}{|C(\bk,\bl|\epsilon)|^2}=1\, ,
\end{align}
for all $\bk$ and $\bl$. Using the asymptotic wavefunction (\ref{e6}) in Eqs.~(\ref{density}) and (\ref{momentum}) one obtains for the asymptotic 
density 
\be\label{asymdens}
\rho_\infty(x,t)\equiv\lim_{t\rightarrow\infty}\rho(x,t)=\frac{m}{t\hbar}\, g\left(\frac{mx}{t\hbar}\right)\, ,
\ee
normalized to $\int \rho_{\infty}(x,t)\, dx=N$ and for the asymptotic momentum distribution
\be\label{asymmom}
n_\infty(p,t)\equiv\lim_{t\rightarrow\infty}n(p,t)=g(p)\, ,
\ee
with a similar normalization and $g(p)$ defined in Eq.~(\ref{defg}).

\begin{figure}[t]
\includegraphics[width=1\linewidth]{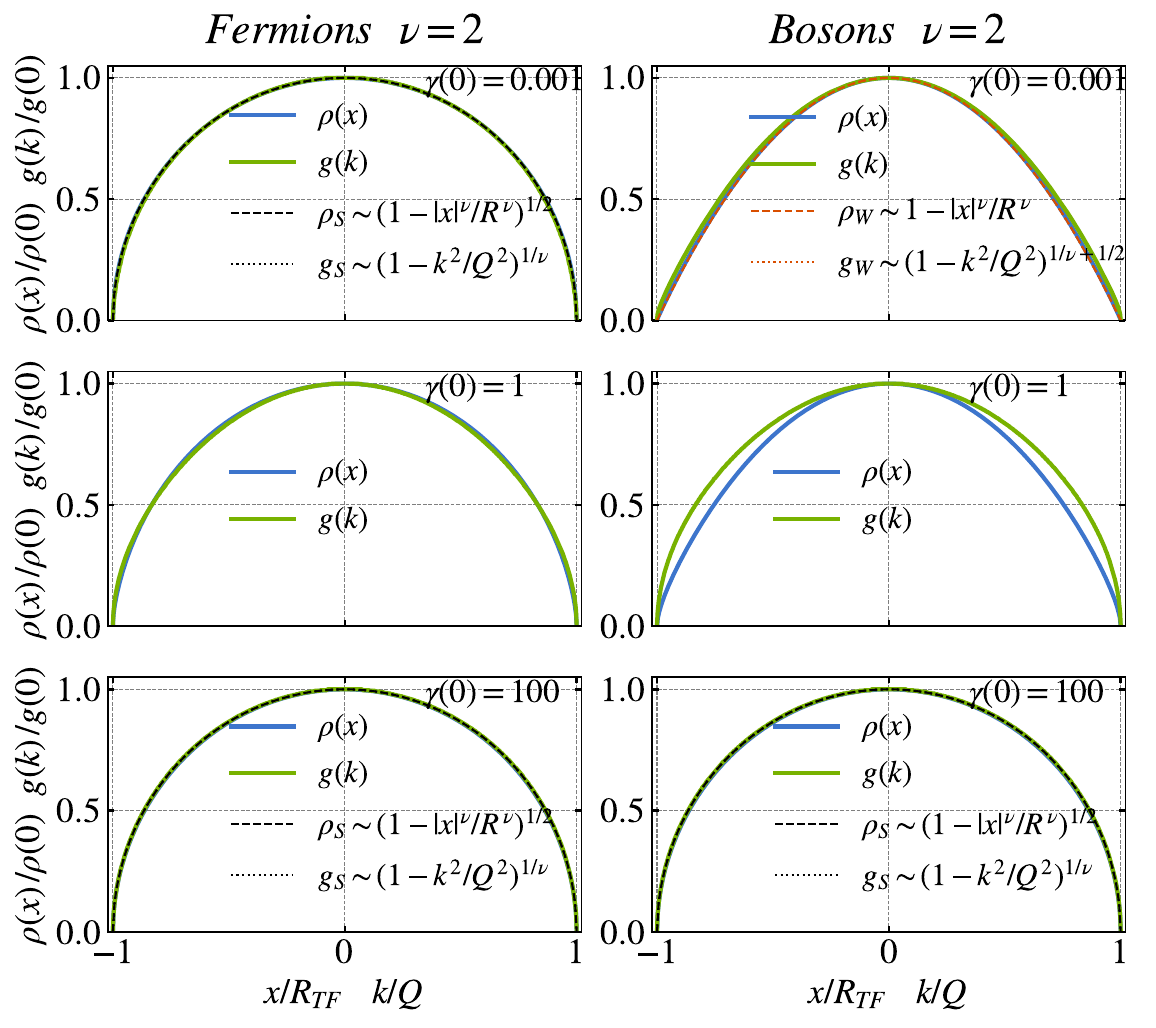}
\caption{Density profile (solid blue line) and quasimomentum distribution (solid green line) of two-component bosonic and fermionic gases in the 
presence of a power-law potential with $\nu=2$ (zero magnetic field).  The asymptotic momentum distribution and density profile can be obtained 
from the quasimomentum distribution of the trapped gas from Eqs.~(\ref{asymdens}) and (\ref{asymmom}).
The first, second, and third rows correspond to $\gamma(0)=0.001$, $\gamma(0)=1$ and $\gamma(0)=100$,  respectively. In the first and third rows, 
the black dashed  (dotted) lines represent the analytical results given by Eq.(\ref{rhos}) [Eq.(\ref{gs})], while the red dashed (dotted) lines 
correspond to the analytical results from  Eq.(\ref{rhow}) [Eq.(\ref{gw})].
}
\label{fig1}
\end{figure}
\begin{figure}[t]
\includegraphics[width=1\linewidth]{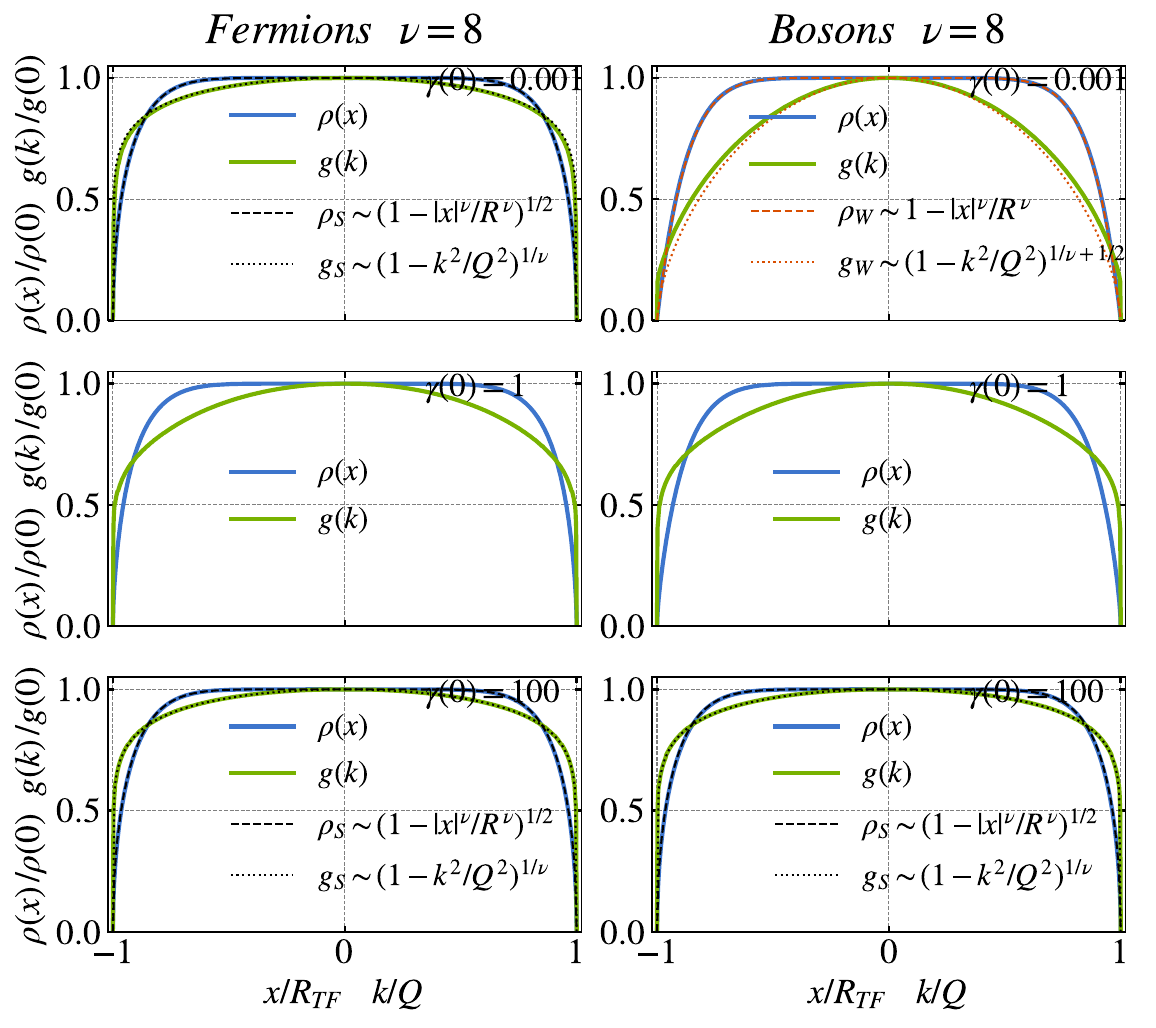}
\caption{Density profile (solid blue line) and quasimomentum distribution (solid green line) of two-component bosonic and fermionic gases in the 
presence of a power-law potential with  $\nu=8$ (zero magnetic field). All other plotted quantities are analogous to those in Fig.~\ref{fig1}.
}
\label{fig2}
\end{figure}

\section{Analytical results}

Eqs.~(\ref{asymdens}) and (\ref{asymmom}) reveal that both the asymptotic density and momentum distribution require the  computation of the 
quasimomentum distribution of the trapped gas, defined in Eq.~(\ref{defg}). For well-behaved trapping potentials of the power-law form, 
$ V(x) = \alpha_\nu |x|^\nu/2 $, it was argued in \cite{CGK15} that this task can be achieved by combining the Bethe ansatz solution for the 
homogeneous system with the Local Density Approximation (LDA) \cite{KGDS05,CGK15}. For a system described by the dimensionless coupling parameter 
$\gamma(0) = \frac{m \mathfrak{g}}{\hbar^2 \rho(0)} = \frac{c}{\rho(0)}$ at the center of the trap (where \(\rho(0)\) is the particle density at 
the center), the LDA dictates that the chemical potential (we consider the case when there is no magnetic field) varies as 
$\mu(x) = \mu(0) - V(x).$
At each point $x$ in the trap, one must compute the local quasimomentum distribution $f(k,x)$ using the integral equations characterizing the 
system at chemical potential $\mu(x)$ \cite{Yang67,Taka99,GBT07,LGSB12}. The quasimomentum distribution is then given by 
$g(k)=\int f(k,x)\, dx,$ while the density profile is given by $ \rho(x) = \int f(k,x)\, dk,$ with the normalization condition $\int f(k,x)\, 
dk\,dx = N.$

In the bosonic case, from a general theorem on Hamiltonians with spin-independent interactions \cite{EL02,YL03}, we know that the ground state 
is fully polarized and is equivalent to the one of the Lieb-Liniger model. Therefore, in the limits of both weak and strong interactions, one 
can use the results derived in \cite{CGK15}. For weak interactions $(\gamma(0) \ll 1)$, the density profile and quasimomentum distribution in the 
trap are given by
\begin{align}
\rho_W(x)&=\rho(0)\left(1-|x|^\nu/R^\nu\right)\Theta(R-|x|)\, ,\label{rhow}\\
g_W(k)   &=g(0)(1-k^2/Q^2)^{1/2+1/\nu}\Theta(Q-|k|)\, ,\label{gw}
\end{align}
where $R = \left( 2\mu(0)/\alpha_\nu \right)^{1/\nu} $ is the Thomas-Fermi radius, $\rho(0)=\mu(0)/\mathfrak{g}$,  and 
$Q = \left( 4 m \mathfrak{g} \rho(0)/\hbar^2 \right)^{1/2} $.
For strong interactions $(\gamma(0) \gg 1)$, the density profile and quasimomentum distribution in the trap are
\begin{align}
\rho_S(x)&=\rho(0)\left(1-|x|^\nu/R^\nu\right)^{1/2}\Theta(R-|x|)\, ,\label{rhos}\\
g_S(k)   &=g(0)(1-k^2/Q^2)^{1/\nu}\Theta(Q-|k|)\, ,\label{gs}
\end{align}
with $\rho(0)=\left[2m\mu(0)/(\hbar^2\pi^2)\right]^{1/2}$, $g(0)=R/\pi$,  and  $Q = \pi \rho(0)$.

In the fermionic case, an interesting phenomenon occurs. In both the weak and strong interaction regimes, the density profile and quasimomentum 
distribution are given by (see Appendix~\ref{as2}) Eqs.~(\ref{rhos}) and (\ref{gs}), with the only difference appearing in $\rho(0)$ and $g(0)$, which differ by a factor of 
2.  In the weakly interacting regime ($\gamma(0) \ll 1$), we have $\rho(0)=2 \left[2m\mu(0)/(\hbar^2\pi^2)\right]^{1/2}$ and $g(0)=2R/\pi$ with 
$Q=\pi\rho(0)/2$, while  for the strongly interacting regime ($\gamma(0) \gg 1$), the expressions become
$\rho(0)=\left[2m\mu(0)/(\hbar^2\pi^2)\right]^{1/2}$ and $g(0)=R/\pi$ with $Q=\pi\rho(0)$.
This behavior arises because, for a given $\mu(0)$, in the limit of weak interactions, the system is almost equivalent to a system of free fermions, 
with two fermions of opposite spins occupying all energy levels up to the Fermi energy. In contrast, in the strong interaction regime, the system 
behaves  like a system of fully polarized fermions due to the additional ``fermionization'' between particles of opposite spins.  The same 
phenomenon occurs in the general case with $\kappa$ components, where the multiplicative factor 2 is replaced by $\kappa$.

In the particular case of a harmonic potential ($\nu=2$ and $\alpha_\nu = m\omega_0^2$), the asymptotic density profile for the Lieb-Liniger model 
exhibits  a self-similar scaling solution of the form  $ \rho(x,t) =\rho(x/b(t))/b(t) $ with $ b(t) = \sqrt{2} \omega_0 t $ for $\gamma(0) \ll 1$ 
and $ b(t) = \omega_0 t $ for $\gamma(0) \gg 1$.  From Eq.~(\ref{asymdens}) and the analytical results presented in this section, one can see that \
both bosonic and fermionic spinor gases exhibit self-similar  density profiles at long times after expansion, but only for $\nu=2$. 
In Fig.~\ref{fig1}, we present numerical results for the initial density profile and quasimomentum distribution for $\gamma(0) = \{0.001,1,100\}$, 
alongside the analytical predictions given by Eqs.~(\ref{rhow}), (\ref{gw}), (\ref{rhos}), and (\ref{gs}). One can also observe that for strong 
interactions and $\nu=2$ the asymptotic momentum distribution retains the same shape as the initial density profile \cite{ASYP21}, a feature that 
remains valid even at finite temperature \cite{Patu23}.  This behavior does not occur for other values of the power-law exponent, as illustrated in 
Fig.~\ref{fig2} for $\nu=8$.

\section{General case}

The analysis for the general case of a system with $\kappa$ components closely follows the computation for the Gaudin-Yang model. In this case, 
the structure of the wavefunctions remains the same as in Eq.~(\ref{e3}), but now $A(Q\bs|P\bk,[\bl])$ represents the coordinate Bethe ansatz 
wavefunctions  of a multicomponent generalization of the inhomogeneous $XXX$ spin chain. The eigenvectors of this inhomogeneous spin chain are 
orthogonal and complete, forming a basis in the spin space with a normalization constant $D(\bk,[\bl])$, which generalizes Eq.~(\ref{e5}). Therefore, 
the large-$t$ analysis and the computation of the asymptotic density and momentum distributions are similar to the two-component case, yielding 
Eqs.~(\ref{asymdens}) and (\ref{asymmom}), with $g(k)$ representing the quasimomentum distribution of the general multicomponent trapped gas.

\section{Conclusions}

In this article, we have analytically demonstrated that, following the release from a trap, the asymptotic density profile and momentum distribution 
of fermionic and bosonic spinor gases with short-range repulsive interactions is determined by the quasimomentum distribution of the trapped gas. 
While our focus has been on power-law traps, which allowed for explicit derivations of the distribution of rapidities in both weak and strong 
interaction regimes, our main result holds independent of the specific external potential. Furthermore, our findings can be readily extended to 
integrable Bose-Fermi mixtures \cite{ID06,FVGM11,DJAR17},  whose Bethe ansatz wavefunctions exhibit a structure similar to Eq.~(\ref{e3}), and to 
systems at finite temperatures. In the latter case, the quasimomentum distribution can be obtained from the Thermodynamic Bethe Ansatz solution 
\cite{Taka99,Taka71,LGB11,CKB09,KC11}, coupled with the local density approximation.
It must be acknowledged that computing the quasimomentum distribution at finite temperature is significantly more challenging than 
in the zero-temperature case. This difficulty arises from the fact that the Thermodynamic Bethe Ansatz introduces an infinite system of coupled 
nonlinear integral equations. In recent years, efficient numerical methods have been developed to address this challenge, enabling the extraction of 
the quasimomentum distribution and other relevant thermodynamic quantities with reasonable accuracy (see \cite{KC11} or Appendix B of \cite{Pozs14}).

\acknowledgments

O.I.P. acknowledges financial support from Grant No. 30N/2023, provided through the National Core Program of the 
Romanian Ministry of Research, Innovation, and Digitization.
G.A.-D. acknowledges financial support from the PRIN 2022 project (2022R35ZBF) – PE2 ``\textit{ManyQLowD – Artificially Devised 
Many-Body Quantum Dynamics in Low Dimensions.}''

\appendix

\begin{widetext}

\section{Derivation of the asymptotic density and momentum for the Gaudin-Yang model}\label{as1}

In this Appendix we present the derivation of the asymptotic density and momentum distribution of the Gaudin-Yang model at large-$t$, following 
release from the confining potential.

\subsection{Properties of the wavefunctions}

In the $[N-M, M]$ sector, where the system consists of $N$ particles, $M$ of which have spin down, the initial trapped state is given by
\be\label{ae1}
|\bm{\phi}_0\rangle = \frac{1}{\sqrt{N!}} \int dx_1 \cdots dx_N \sum_{\{\sigma_1, \cdots, \sigma_N\}\in\{\uparrow,\downarrow\}}^{[N-M, M]} \phi_0(\bm{x}; \bs) 
\Psi_{\sigma_N}^\dagger(x_N) \cdots \Psi_{\sigma_1}^\dagger(x_1) |0\rangle,
\ee
where $\bm{x} = (x_1, \cdots, x_N)$ and $\bs = (\sigma_1, \cdots, \sigma_N)$. The integrable Gaudin-Yang eigenstates are defined as
\be\label{ae2}
|\bm{\psi}(\bm{k}, \bm{\lambda})\rangle = \frac{1}{\sqrt{N!}} \int dx_1 \cdots dx_N \sum_{\{\sigma_1, \cdots, \sigma_N\}\in\{\uparrow,\downarrow\}}^{[N-M, M]} 
\psi(\bm{x}; \bs | \bm{k}, \bm{\lambda}) \Psi_{\sigma_N}^\dagger(x_N) \cdots \Psi_{\sigma_1}^\dagger(x_1) |0\rangle,
\ee
where $\bm{k} = (k_1, \cdots, k_N)$ are the quasimomenta and $\bm{\lambda} = (\lambda_1, \cdots, \lambda_M)$ are the spin rapidities.
The wavefunction $\psi(\bm{x}; \bs | \bm{k}, \bm{\lambda})$ is explicitly given by (see the main text for the definitions of the functions) 
\cite{POK12,EFGKK05}
\be\label{ae3}
\psi(\bm{x}; \bs | \bm{k}, \bm{\lambda}) = \frac{1}{C(\bm{k}, \bm{\lambda} | \epsilon)} \sum_{Q \in S_N} \Theta(Q \bm{x}) (\epsilon)^Q 
\sum_{P \in S_N} (-1)^P \varphi(P\bm{k} | \epsilon) A(Q \bs | P\bm{k}, \bm{\lambda}) e^{i (P\bm{k}, Q\bm{x})},
\ee
where $C(\bm{k}, \bm{\lambda} | \epsilon)$ is the normalization constant and $\epsilon = +1$ for bosons or $\epsilon = -1$ for fermions.
Let us prove some of the properties of the Gaudin-Yang wavefunctions.

\textbf{1. Spin rapidity permutation invariance}
\be\label{iden1}
\psi(\bm{x}; \bs | \bm{k}, V' \bm{\lambda}) = \psi(\bm{x}; \bs | \bm{k}, \bm{\lambda}) \quad \text{for all} \quad V' \in S_M.
\ee
\textit{Proof.} Since $C(\bm{k}, \bm{\lambda} | \epsilon)$ is symmetric in both $\bm{k}$ and $\bm{\lambda}$, it suffices to check the dependence 
on $\bm{\lambda}$ in $A(Q \bs | P\bm{k}, \bm{\lambda})$:
\[
A(Q \bs | P\bm{k}, V'\bm{\lambda}) = \sum_{R \in S_M} B(RV' \bm{\lambda}) \prod_{l=1}^M F_{P\bm{k}}(\lambda_{(RV')_l}; y_l).
\]
Defining $R' = R V'$, we obtain
\[
A(Q \bs | P\bm{k}, V'\bm{\lambda}) = \sum_{R' \in S_M} B(R' \bm{\lambda}) \prod_{l=1}^M F_{P\bm{k}}(\lambda_{R'_l}; y_l) = A(Q \bs | 
P\bm{k}, \bm{\lambda}).
\]
The symmetry of $C(\bm{k}, \bm{\lambda} | \epsilon)$ completes the proof.

\textbf{2. Antisymmetry under quasimomentum permutations}
\be\label{iden2}
\psi(\bm{x}; \bs | V\bm{k}, \bm{\lambda}) = (-1)^V \psi(\bm{x}; \bs | \bm{k}, \bm{\lambda}) \quad \text{for all} \quad V \in S_N.
\ee
\textit{Proof.} Starting from the definition
\begin{align*}
\psi(\bm{x}; \bs | V\bm{k}, \bm{\lambda}) &= \frac{1}{C(V\bm{k}, \bm{\lambda} | \epsilon)} \sum_{Q \in S_N} \Theta(Q\bm{x}) (\epsilon)^Q 
\sum_{P \in S_N} (-1)^P \varphi(P V\bm{k} | \epsilon) A(Q \bs | P V\bm{k}, \bm{\lambda}) e^{i (P V\bm{k}, Q\bm{x})}\, , \\
&= \frac{1}{C(\bm{k}, \bm{\lambda} | \epsilon)} \sum_{Q \in S_N} \Theta(Q\bm{x}) (\epsilon)^Q \sum_{P' \in S_N} (-1)^{P'} (-1)^V 
\varphi(P'\bm{k} | \epsilon) A(Q \bs | P'\bm{k}, \bm{\lambda}) e^{i (P'\bm{k}, Q\bm{x})}\, , \\
&= (-1)^V \psi(\bm{x}; \bs | \bm{k}, \bm{\lambda}),
\end{align*}
where in the second line we have used $P' = P V$ and $(-1)^{P'} = (-1)^P (-1)^V$.

\textbf{3. Coordinate and spin permutation symmetry}
\be\label{iden3}
\psi(V\bm{x}; V\bs | \bm{k}, \bm{\lambda}) = (\epsilon)^V \psi(\bm{x}; \bs | \bm{k}, \bm{\lambda}) \quad \text{for all} \quad V \in S_N.
\ee
\textit{Proof.} Applying the permutation $V$ to both $\bm{x}$ and $\bs$
\begin{align*}
\psi(V\bm{x}; V\bs | \bm{k}, \bm{\lambda}) &= \frac{1}{C(\bm{k}, \bm{\lambda} | \epsilon)} \sum_{Q \in S_N} \Theta(Q V\bm{x}) (\epsilon)^Q 
\sum_{P \in S_N} (-1)^P \varphi(\bm{k} | \epsilon) A(Q V\bs | P\bm{k}, \bm{\lambda}) e^{i (P\bm{k}, Q V\bm{x})}\, , \\
&= \frac{\epsilon^V}{C(\bm{k}, \bm{\lambda} | \epsilon)} \sum_{Q' \in S_N} \Theta(Q' \bm{x}) (\epsilon)^{Q'} \sum_{P' \in S_N} (-1)^{P'} 
\varphi(\bm{k} | \epsilon) A(Q' \bs | P'\bm{k}, \bm{\lambda}) e^{i (P'\bm{k}, Q' \bm{x})}\, , \\
&= (\epsilon)^V \psi(\bm{x}; \bs | \bm{k}, \bm{\lambda}),
\end{align*}
where in the second line we have introduced $Q'=QV$ with $(\epsilon)^{Q'}=(\epsilon)^Q (\epsilon)^V$.

\subsection{Expansion of the initial state}

In the $[N-M,M]$ sector the expansion of the initial state (\ref{ae1}) in terms of the Gaudin-Yang 
eigenstates takes the form 
\be\label{ae4}
|\boldsymbol{\phi}_0\rangle=\int_{k_1<\cdots<k_N}dk_1\cdots dk_N\sum_{\lambda_1<\cdots<\lambda_M}b(\bk,\bl)|\boldsymbol{\psi}(\bk,\bl)\rangle\, ,
\ee
where we have used the resolution of identity $\boldsymbol{1}=\int_{k_1<\cdots<k_N}\sum_{\lambda_1<\cdots<\lambda_M}|\boldsymbol{\psi}(\bk,\bl)
\rangle\langle \boldsymbol{\psi}(\bk,\bl)|$. The expansion coefficients $b(\bk,\bl)$ are defined as 
\be
b(\bk,\bl)\equiv \langle\boldsymbol{\psi}(\bk,\bl)|\boldsymbol{\phi}_0\rangle=
\int dx_1\cdots dx_N \sum_{\sigma_1,\cdots,\sigma_N=\{\uparrow,\downarrow\}}^{[N-M,M]}\psi^*(\bx;\bs|\bk,\bl)\phi_0(\bx;\bs)\, .
\ee
Using the symmetry identities (\ref{iden2}) and (\ref{iden3}) it follows that the expansion coefficients satisfy 
\begin{align}
b(V\bk,\bl)&=(-1)^Vb(\bk,\bl)\, ,\ \ \mbox{ for all } V\in S_N\, ,\label{bsymk}\\
b(\bk,V'\bl)&=b(\bk,\bl)\, ,\ \ \ \ \qquad \mbox{ for all } V'\in S_M\label{bsyml}\, .
\end{align}
These properties imply that the product $b(\bk,\bl)|\boldsymbol{\psi}(\bk,\bl)\rangle$ is symmetric in both $\bk$ and $\bl$. As a result the 
integration and summation in Eq.~(\ref{ae4}) can be extended over the entire space   by dividing with $N!M!$ ($b(\bk,\bl)=0$ when two $k$'s or 
two $\lambda$'s are equal). This yields 
\be\label{ae5}
|\boldsymbol{\phi}_0\rangle=\frac{1}{N!}\frac{1}{M!}\int dk_1\cdots dk_N\sum_{\lambda_1}\cdots\sum_{\lambda_M} b(\bk,\bl)|\boldsymbol{\psi}
(\bk,\bl)\rangle\, .
\ee
Using the normalization of the initial state $\langle \boldsymbol{\phi}_0    |\boldsymbol{\phi}_0\rangle=1$, and  the orthonormality of the 
Gaudin-Yang eigenstates,
$\langle\boldsymbol{\psi}(\bk',\bl') |\boldsymbol{\psi}(\bk,\bl)\rangle=\delta_{\bk,\bk'} \delta_{\bl,\bl'}$, we find
\begin{align}
1&=\int_{k_1<\cdots<k_N}dk_1\cdots dk_N\sum_{\lambda_1<\cdots<\lambda_M}|b(\bk,\bl)|^2\, ,\\
&=\frac{1}{N!}\frac{1}{M!}\int dk_1\cdots dk_N\sum_{\lambda_1}\cdots\sum_{\lambda_M} |b(\bk,\bl)|^2\, .\label{normb}
\end{align}

\subsection{Asymptotic wavefunctions}

For $t>0$ the time evolution is dictated by the final Hamiltonian $H_F$. Using the relation  $e^{- i tH_F/\hbar}|\boldsymbol{\psi}(\bk,\bl)
\rangle=e^{-i t\sum_{j=1}^N\hbar k_j^2/2m }|\boldsymbol{\psi}(\bk,\bl)\rangle$
and substituting in Eq.~(\ref{ae5}) we obtain
\be
|\boldsymbol{\phi}_0(t)\rangle\equiv e^{- i tH_F/\hbar}|\boldsymbol{\phi}_0\rangle=\frac{1}{N!}\frac{1}{M!}\int dk_1\cdots dk_N\sum_{\lambda_1}
\cdots\sum_{\lambda_M} b(\bk,\bl)
e^{-i t\sum_{j=1}^N\hbar k_j^2/2m }|\boldsymbol{\psi}(\bk,\bl)\rangle\, .
\ee
At the wavefunction level this translates to
\be
\phi_0(\bx;\bs|t)=\frac{1}{N!}\frac{1}{M!}\int dk_1\cdots dk_N\sum_{\lambda_1}\cdots\sum_{\lambda_M} b(\bk,\bl)\psi(\bx;\bs|\bk,\bl)
e^{-i t\sum_{j=1}^N\hbar k_j^2/2m }\, .
\ee
Substituting the explicit form of the Gaudin-Yang wavefunctions and using the symmetry property $b(\bk,\bl)=(-1)^Pb(P\bk,\bl)$ for any 
$P\in S_N$ [see Eq.~(\ref{bsymk})]
we find 
\begin{align}
\phi_0(\bx;\bs|t)&=\frac{1}{N!}\frac{1}{M!}\int dk_1\cdots dk_N\sum_{\lambda_1}\cdots\sum_{\lambda_M} \sum_{Q\in S_N}\Theta(Q\bx)(\epsilon)^Q\nonumber\\
&\qquad\qquad\qquad\qquad\times\sum_{P\in S_N}\underbrace{\frac{b(P\bk,\bl)}{C(\bk,\bl|\epsilon)}\varphi(P\bk|\epsilon)A(Q\bs|P\bk,\bl)}_{H(\bk,\bl|\epsilon)}
 e^{i\sum_{j=1}^N\left(k_{P_j}x_{Q_j}-\frac{\hbar t}{2m}k_{P_j}^2\right)}\, .
\end{align}
Here, $H(\bk,\bl|\epsilon)$ is a slowly varying function of $\bk$ with the time dependence contained in the exponential. Since we are interested in 
the large-time behavior of the wavefunction, we apply the stationary phase approximation to the $N$-dimensional integral. The general stationary phase 
formula is \cite{BH86}:
\begin{align}
\lim_{t\rightarrow\infty}\int h(k) e^{i t f(k)}\, dk\sim \left(\frac{2\pi}{t|f''(k_0)|}\right)^{1/2}e^{i\, \mbox{sign}\,f''(k_0)\pi/4}
h(k_0)e^{i tf(k_0)}\, ,
\end{align}
where $k_0$ is the stationary point determined by  $f'(k_0)=0$. In our case $f(k)=\frac{k x}{t}-\frac{\hbar k^2}{2m}$ yielding the stationary point 
$k_0=\frac{m x}{t\hbar}$.
Performing this analysis for each integral, we obtain the asymptotic form of the wavefunction
\begin{align}\label{ae6}
\phi_0(\bx;\bs|t)=\frac{1}{M!}\left(\frac{2\pi m}{t\hbar}\right)^{\frac{N}{2}}\sum_{\lambda_1}\cdots\sum_{\lambda_{M}}&\sum_{Q\in S_N}\Theta(Q\bx)
\frac{b(Q\tilde{\bx},\bl)}{C(Q\tilde{\bx},\bl|\epsilon)}\varphi(Q\tilde{\bx}|\epsilon)A(Q\bs|Q\tilde{\bx},\bl)
 e^{-i N\pi/4}e^{i\sum_{j=1}^N m x_{Q_j}^2/2m}\, ,
\end{align}
where the rescaled coordinates are defined as $Q\tilde{\bx}=\left(\frac{ m x_{Q_1}}{t\hbar}, \cdots, \frac{ m x_{Q_N}}{t\hbar}\right)$.

\subsection{Normalization of the asymptotic wavefunctions}

Since we have assumed that the initial state is normalized, the unitarity of time evolution ensures that the asymptotic state remains normalized.
Therefore, we have 
\begin{align}
1&=\int dx_1\cdots dx_N\sum_{\sigma_1,\cdots,\sigma_N=\{\uparrow,\downarrow\}}^{[N-M,M]}\phi_0^*(\bx;\bs|t=0)\phi_0(\bx;\bs|t=0)\, ,\nonumber\\
&=\int dx_1\cdots dx_N\sum_{\sigma_1,\cdots,\sigma_N=\{\uparrow,\downarrow\}}^{[N-M,M]}\phi_0^*(\bx;\bs|t)\phi_0(\bx;\bs|t)\, .
\end{align}
Substituting the asymptotic form of the wavefunction from (\ref{ae6})  into the normalization condition, we find
\begin{align}
1&=\frac{1}{(M!)^2}\left(\frac{2\pi m}{t\hbar}\right)^{N}\int dx_1\cdots dx_N\sum_{\sigma_1,\cdots,\sigma_N=\{\uparrow,\downarrow\}}^{[N-M,M]}\nonumber\\
&\qquad\qquad\times \sum_{\lambda_1'}\cdots\sum_{\lambda_{M}'}\sum_{Q'\in S_N}\Theta(Q'\bx)
\frac{b^*(Q'\tilde{\bx},\bl)}{C^*(Q'\tilde{\bx},\bl'|\epsilon)}\varphi^*(Q'\tilde{\bx}|\epsilon)A^*(Q'\bs|Q'\tilde{\bx},\bl')\nonumber\\
&\qquad\qquad\times \sum_{\lambda_1}\cdots\sum_{\lambda_{M}}\sum_{Q\in S_N}\ \Theta(Q\bx)
\ \frac{b(Q\tilde{\bx},\bl)}{C(Q\tilde{\bx},\bl|\epsilon)}\ \ \varphi(Q\tilde{\bx}|\epsilon)A(Q\bs|Q\tilde{\bx},\bl)\, .
\end{align}
This rather involved expression simplifies significantly by using two key observations: a) the product of Heaviside functions reduces as 
$\Theta(Q'\bx)\Theta(Q\bx)=\delta_{Q,Q'}\Theta(Q\bx)$ and b) the orthogonality of the spin-dependent part 
\begin{align}\label{ae7}
\sum_{\sigma_1,\cdots,\sigma_N=\{\uparrow,\downarrow\}}^{[N-M,M]}
A^*(Q\bs|P\bk,\bl')A(Q\bs|P\bk,\bl)=\delta_{\bl,\bl'} D(\bk,\bl)\, ,
\end{align}
where $D(\bk,\bl)$ is symmetric in both $\bk$ and $\bl$. Implementing these simplifications, the normalization condition becomes
\begin{align}
1&=\frac{1}{(M!)^2}\left(\frac{2\pi m}{t\hbar}\right)^{N}\int dx_1\cdots dx_N
\sum_{\lambda_1}\cdots\sum_{\lambda_{M}}\sum_{Q\in S_N}\ \Theta(Q\bx)
 \frac{|b(Q\tilde{\bx},\bl)|^2}{|C(Q\tilde{\bx},\bl|\epsilon)|^2}|\varphi(Q\tilde{\bx}|\epsilon)|^2 |D(Q\tilde{\bx},\bl)|^2\, .
\end{align}
Next, we observe that the combination of functions after the generalized Heaviside function is fully symmetric in $\tilde{\bx}$ and can be pulled 
outside  the sum over permutations.  Using the identity  $\sum_{Q\in S_N}\Theta(Q\bx)=\boldsymbol{1}_{\mathbb{R}^N}$ we obtain 
\begin{align}
1&=\frac{1}{(M!)^2}\left(\frac{2\pi m}{t\hbar}\right)^{N}\int dx_1\cdots dx_N
\sum_{\lambda_1}\cdots\sum_{\lambda_{M}} \frac{|b(\tilde{\bx},\bl)|^2}{|C(\tilde{\bx},\bl|\epsilon)|^2}|\varphi(\tilde{\bx}|\epsilon)|^2 
|D(\tilde{\bx},\bl)|^2\, ,\nonumber\\
1&=\frac{(2\pi)^N}{(M!)^2}\int dx_1\cdots dx_N
\sum_{\lambda_1}\cdots\sum_{\lambda_{M}} \frac{|b(\bk,\bl)|^2}{|C(\bk,\bl|\epsilon)|^2}|\varphi(\bk|\epsilon)|^2 |D(\bk,\bl)|^2\, ,
\end{align}
where the simplified second line was obtained after the change of variables $mx_j/t\hbar=k_j$ for $j=1,\cdots,N$. Finally, recalling the 
normalization of the expansion coefficients  $b(\bk,\bl)$ from Eq.~(\ref{normb}), we deduce that the above relation holds  if the 
following condition is satisfied
\begin{align}\label{cond}
(2\pi)^N \frac{N!}{M!}\frac{|\varphi (\bk|\epsilon)|^2 |D(\bk,\bl)|^2}{|C(\bk,\bl|\epsilon)|^2}=1\, ,
\end{align}
for all values of $\bk$ and $\bl$. 

\subsection{Asymptotic density}

Now we are ready to compute one of our main quantities of interest: the total density. Expressed in terms of the wavefunctions, the density reads
\begin{align}
\rho(x,t)=N\sum_{\sigma_1,\cdots,\sigma_N=\{\uparrow,\downarrow\}}^{[N-M,M]}\int dx_2\cdots dx_N
\phi_0^*(x,x_2,\cdots,x_N;\bs|t)\phi_0(x,x_2,\cdots,x_N;\bs|t)\, .
\end{align}
Introducing the shorthand notation $\bar{\bx}=(x,x_2,\cdots,x_N)$, and the rescaled coordinates  $Q\tilde{\bar{\bx}}=(\frac{m  x_{Q_1}}{t\hbar}, 
\frac{m x_{Q_2}}{t\hbar},\cdots,\frac{m  x_{Q_N}}{t\hbar})$ with $x_1=x$ we substitute the asymptotic expression for the wavefunction from 
Eq.~(\ref{ae6}) into the density expression. This yields the following form for the asymptotic density $\rho_\infty(x,t)\equiv\lim_{t\rightarrow\infty}
\rho(x,t)$ the following expression
\begin{align}\label{asyfor}
\rho_\infty(x,t)&=\frac{N}{(M!)^2}\left(\frac{2\pi m}{t\hbar}\right)^N \sum_{\sigma_1,\cdots,\sigma_N=\{\uparrow,\downarrow\}}^{[N-M,M]}\int dx_2\cdots dx_N\nonumber\\
&\qquad\times\sum_{\lambda_1'}\cdots\sum_{\lambda_M'}\sum_{Q'\in S_N}\Theta(Q'\bar{\bx})(\epsilon)^{Q'}
\frac{b^*\left(Q'\tilde{\bar{\bx}},\bl'\right)}{C^*\left(Q'\tilde{\bar{\bx}},\bl'|\epsilon\right)}
\varphi^*\left(Q'\tilde{\bar{\bx}}|\epsilon\right) A^*\left(Q'\bs|Q'\tilde{\bar{\bx}},\bl'\right)\nonumber\\
&\qquad\times\sum_{\lambda_1}\cdots\sum_{\lambda_M}\sum_{Q\in S_N}\Theta(Q\bar{\bx})(\epsilon)^{Q}
\frac{b\left(Q\tilde{\bar{\bx}},\bl\right)}{C\left(Q\tilde{\bar{\bx}},\bl|\epsilon\right)}
\varphi\left(Q\tilde{\bar{\bx}}|\epsilon\right) A\left(Q\bs|Q\tilde{\bar{\bx}},\bl\right)\, .
\end{align}
This expression can be simplified by exploiting the identity $\Theta(Q'\bx)\Theta(Q\bx)=\delta_{Q,Q'}\Theta(Q\bx)$  and the orthogonality property of 
$A(Q\bs | Q\tilde{\bar{\bx}}, \bl)$, given by Eq.~(\ref{ae7}). Using this, the asymptotic density reduces to 
\begin{align}\label{ae8}
\rho_\infty(x,t)&=\frac{N}{(M!)^2}\left(\frac{2\pi m}{t\hbar}\right)^N \int dx_2\cdots dx_N\\
&\qquad\times\sum_{\lambda_1}\cdots\sum_{\lambda_M}\sum_{Q\in S_N}\Theta(Q\bar{\bx})
\frac{|b\left(Q\tilde{\bar{\bx}},\bl\right)|^2}{|C\left(Q\tilde{\bar{\bx}},\bl|\epsilon\right)|2}
|\varphi\left(Q\tilde{\bar{\bx}}|\epsilon\right)|^2 |D\left(Q\tilde{\bar{\bx}},\bl\right)|^2\, .
\end{align}
Now, defining the function
\be
H(Q\tilde{\bar{\bx}},\bl|\epsilon)=\frac{|b\left(Q\tilde{\bar{\bx}},\bl\right)|^2}{|C\left(Q\tilde{\bar{\bx}},\bl|\epsilon\right)|2}
|\varphi\left(Q\tilde{\bar{\bx}}|\epsilon\right)|^2 |D\left(Q\tilde{\bar{\bx}},\bl\right)|^2\, ,
\ee
which is symmetric in both $\bx$ and $\bl$ we observe that $H(Q\tilde{\bar{\bx}},\bl|\epsilon)=H(\bar{\bx},\bl|\epsilon)$. As a result, $H$ can be moved 
outside the sum over $Q$, and we can use the identity $\sum_{Q\in S_N}\Theta(\bar{\bx})=\boldsymbol{1}_{\mathbb{R}^{N-1}}$ to find
\begin{align}\label{ae8b}
\rho_\infty(x,t)&=\frac{N}{(M!)^2}\left(\frac{2\pi m}{t\hbar}\right)^N \int dx_2\cdots dx_N\sum_{\lambda_1}\cdots\sum_{\lambda_M}H(\bar{\bx},\bl|\epsilon)\, ,\\
&=\frac{N}{(M!)^2}(2\pi)^N\frac{m}{t\hbar}\int dk_2\cdots dk_N\sum_{\lambda_1}\cdots\sum_{\lambda_M}H(\bar{\bk},\bl|\epsilon)\, ,
\end{align}
where, in the second line, we changed variables to $k_j=mx_j/th$ for $j=2,\cdots,N$ and introduced the notation 
$\bar{\bk}=\left(mx/t\hbar,k_2,\dots,k_N\right)$. From Eq.~(\ref{cond}), we know that 
\begin{align}
(2\pi)^N \frac{N!}{M!}\frac{|\varphi (\bar{\bk}|\epsilon)|^2 |D(\bar{\bk},\bl)|^2}{|C(\bar{\bk},\bl|\epsilon)|^2}=1\, ,
\end{align}
which, upon substitution into Eq.~(\ref{ae8}), leads to the main result
\begin{align}
\rho_\infty(x,t)&=\frac{m}{t\hbar} \frac{N}{N!}\frac{1}{M!}\int dk_2\cdots dk_N\sum_{\lambda_1}\cdots\sum_{\lambda_M}|
b\left(\frac{mx}{t\hbar},k_2,\cdots,k_N,\bl\right)|^2\, ,\\
\rho_\infty(x,t)&=\frac{m}{t\hbar}\,g\left(\frac{mx}{t\hbar}\right)\, ,\label{maindens}
\end{align}
where $g(k)$ is the quasimomentum distribution of the trapped gas, defined in Eq.~(\ref{defg}).

\subsection{Aysmptotic momentum distribution}

The momentum distribution is obtained by taking the Fourier transform of the field-field correlator
\be\label{ae9}
n(p,t)=\frac{1}{2\pi}\int e^{i p(x-y)}\rho(x,y|t)\, dxdy\, ,
\ee
where the field-field correlator  is defined as
\begin{align}
\rho(x,y|t)=N\sum_{\sigma_1,\cdots,\sigma_N=\{\uparrow,\downarrow\}}^{[N-M,M]}\int dx_2\cdots dx_N
\phi_0^*(x,x_2,\cdots,x_N;\bs|t)\phi_0(y,x_2,\cdots,x_N;\bs|t)\, .
\end{align}
It is important to note that, when evaluated at $x = y$, the field-field correlator reduces to the density: $\rho(x, x | t) = \rho(x, t)$, as studied in the previous section.
The asymptotic momentum distribution is defined as $n_\infty(p,t)=\lim_{t\rightarrow\infty}n(p,t)$ with $\rho_\infty(x,y|t)=\lim_{t\rightarrow\infty}\rho(x,y|t)$
in Eq.~(\ref{ae9}). Using the asymptotic form of the wavefunction (\ref{ae6}), the asymptotic field-field correlator can be expressed as
\begin{align}
\rho_\infty(x,y|t)&=e^{i \frac{m}{2t\hbar}(x^2-y^2)} \tilde{\rho}_\infty(x,y|t)\, ,
\end{align}
where $\tilde{\rho}_\infty(x,y|t)$ a slowly varying function of $x$ and $y$. Notably, it reduces to the density when $x = y$, i.e., $\tilde{\rho}_\infty(x, x | t) = 
\rho_\infty(x, t)$, as seen in Eq.~(\ref{asyfor}). Substituting this  form  in Eq.~(\ref{ae9}), the large-$t$ limit can be evaluated using the stationary phase approximation, 
yielding
\be
n_\infty(p,t)=\frac{1}{2\pi}\left(\frac{2\pi t\hbar}{m}\right)\tilde{\rho}_\infty\left( \frac{p\hbar t}{m},\frac{p\hbar t}{m}|t \right)\, .
\ee
Since $\tilde{\rho}_\infty(x,x|t)=\rho_\infty(x,t)$, and using the result from Eq.~(\ref{maindens}), we obtain the final expression for the asymptotic momentum distribution
\be
n_\infty(p,t)=g(p)\, .
\ee

\section{Derivation of the density profile and quasimomentum distribution for the fermionic Gaudin-Yang model}\label{as2}

The ground state of the fermionic Gaudin-Yang model in the absence of a magnetic field is described by the following system of integral equations \cite{Yang67,Taka99}
\begin{align}
\rho_c(k)&=\frac{1}{2\pi}+\int_{-\infty}^\infty \frac{2c}{\pi[c^2+4(k-\lambda)^2]}\rho_s(\lambda)\, d\lambda\, ,\\
\rho_s(\lambda)&=\frac{1}{2c}\int_{-Q}^{Q}\sech\left[\frac{\pi(k-\lambda)}{c}\right]\rho_c(k)\, dk\, ,
\end{align}
where $Q$ is the maximum quasimomentum.  The energy density, total density of particles and density of particles with spin down are given by
\be
e\equiv\frac{E}{L}=\frac{\hbar^2}{2m}\int_{-Q}^{Q}k^2\rho_c(k)\, dk\, ,\ \ \
\rho\equiv \frac{N}{L}=\int_{-Q}^{Q}\rho_c(k)\, dk\, , \ \ \
\rho_\downarrow\equiv \frac{M}{L}=\frac{\rho}{2}\, .
\ee

\subsection{Strong interacting regime $\gamma\gg 1$}

In the strong interacting regime, $\gamma=c/n\gg 1$, the quasimomentum distribution simplifies to 
\be
\rho_c(k)=\frac{1}{2\pi}\Theta(Q-|k|)\, ,
\ee
leading to the following expressions for the density and energy
\be
\rho=\frac{Q}{\pi}\, ,\ \ \ e=\frac{\hbar^2}{2m}\frac{Q^3}{3\pi}\, .
\ee
Using the definition of the chemical potential,  $\mu=\6 E/\6 N=\6(eL)/\6 N$, we derive the equation of state 
\be
\mu=\frac{\hbar^2}{2m}\pi^2\rho^2\, \ \ \Longrightarrow \rho=\left(\frac{2 m\mu}{\pi^2\hbar^2}\right)^{1/2}\, .
\ee
These results hold for a homogeneous system. In the presence of a power-law trapping potential, $V(x)=\alpha_\nu |x|^\nu/2$, the local density approximation 
(LDA) gives a position-dependent chemical potential $\mu(x)=\mu(0)-V(x)$ where $\mu(0)$ the chemical potential at the trap center.
Consequently, both the density and the maximum quasimomentum become position dependent. It is common to introduce the dimensionless interaction parameter at 
the trap center, $\gamma(0)=c/\rho(0)$. For the trapped system, the local quasimomentum distribution function is
is 
\be\label{ae11}
f(k,x)=\frac{1}{2\pi}\Theta(Q(x)-|k|)=\frac{1}{2\pi}\Theta(\pi \rho(x)-|k|)\, .
\ee
The density profile is determined from the position dependent equation of state $\rho(x)=\left(\frac{2 m\mu(x)}{\pi^2\hbar^2}\right)^{1/2}$ with 
 $\mu(x)=\mu(0)-V(x)$ yielding 
\be\label{ae12}
\rho(x)=\rho(0)\left(1-|x|^\nu/R^\nu\right)^{1/2}\Theta(R-|x|)\, ,\ \ \ \rho(0)=\left(\frac{2 m\mu(0)}{\pi^2\hbar^2}\right)^{1/2}\, ,
\ee
where the Thomas-Fermi radius  $R=(2\mu(0)/\alpha_\nu)^{1/2}$ is set by the condition $\mu(R)=0$.  
The quasimomentum distribution is then obtained by integrating over position, $g(k)=\int f(k,x)\, dx$,  with the result
\be\label{ae13}
g(k)=\frac{R}{\pi}\left[1-(k/Q)^2\right]^{1/\nu}\Theta(Q-|k|)\, ,\ \ \ Q=\pi\rho(0)\, .
\ee

\subsection{Weak interacting regime $\gamma\ll 1$}

In the weak interacting regime, $\gamma\ll 1$, the quasimomentum density takes the form (see Chap 2.2.7. of \cite{Taka99})
\be
\rho_c(k)=\frac{1}{\pi}\Theta(Q-|k|)\, ,
\ee
leading to
\be
\rho=\frac{2 Q}{\pi}\, ,\ \ \ e=\frac{\hbar^2}{2m}\frac{2 Q^3}{3\pi}\, .
\ee
The equation of state for the homogeneous system becomes
\be
\rho=2 \left(\frac{2 m\mu}{\pi^2\hbar^2}\right)^{1/2}\, .
\ee
In the presence of the trapping potential the local quasimomentum distribution  function $f(k,x)$ is 
\be\label{ae14}
f(k,x)=\frac{1}{\pi}\Theta(Q(x)-|k|)=\frac{1}{2\pi}\Theta(\pi \rho(x)-|k|)\, .
\ee
and the density profile becomes
\be\label{ae15}
\rho(x)=\rho(0)\left(1-|x|^\nu/R^\nu\right)^{1/2}\Theta(R-|x|)\, ,\ \ \ \rho(0)=2\left(\frac{2 m\mu(0)}{\pi^2\hbar^2}\right)^{1/2}\, .
\ee
The quasimomentum distribution is
\be\label{ae16}
g(k)=\frac{2 R}{\pi}\left[1-(k/Q)^2\right]^{1/\nu}\Theta(Q-|k|)\, ,\ \ \ Q=\pi\rho(0)/2\, .
\ee
It is worth noting that in the weak interaction regime, the density profile and quasimomentum distribution [Eqs.~(\ref{ae15}) and (\ref{ae16})] are exactly 
twice those obtained in the strong interaction regime [Eqs.~(\ref{ae12}) and (\ref{ae13})].

\end{widetext}

%

\end{document}